\documentstyle[12pt]{article}
\expandafter\ifx\csname amssym.def\endcsname\relax \else\endinput\fi
\expandafter\edef\csname amssym.def\endcsname{%
       \catcode`\noexpand\@=\the\catcode`\@\space}
\catcode`\@=11
\def\undefine#1{\let#1\undefined}
\def\newsymbol#1#2#3#4#5{\let\next@\relax
 \ifnum#2=\@ne\let\next@\msafam@\else
 \ifnum#2=\tw@\let\next@\msbfam@\fi\fi
 \mathchardef#1="#3\next@#4#5}
\def\mathhexbox@#1#2#3{\relax
 \ifmmode\mathpalette{}{\m@th\mathchar"#1#2#3}%
 \else\leavevmode\hbox{$\m@th\mathchar"#1#2#3$}\fi}
\def\hexnumber@#1{\ifcase#1 0\or 1\or 2\or 3\or 4\or 5\or 6\or 7\or 8\or
 9\or A\or B\or C\or D\or E\or F\fi}

\ifcase\@ptsize
  \font\tenmsa=msam10
  \font\sevenmsa=msam7
  \font\fivemsa=msam5
\or
  \font\tenmsa=msam10  scaled \magstephalf
  \font\sevenmsa=msam7 scaled \magstephalf
  \font\fivemsa=msam5  scaled \magstephalf
\or
  \font\tenmsa=msam10  scaled \magstep1
  \font\sevenmsa=msam7 scaled \magstep1
  \font\fivemsa=msam5  scaled \magstep1
\fi

\newfam\msafam
\textfont\msafam=\tenmsa
\scriptfont\msafam=\sevenmsa
\scriptscriptfont\msafam=\fivemsa
\edef\msafam@{\hexnumber@\msafam}
\mathchardef\dabar@"0\msafam@39
\def\dashrightarrow{\mathrel{\dabar@\dabar@\mathchar"0\msafam@4B}}
\def\dashleftarrow{\mathrel{\mathchar"0\msafam@4C\dabar@\dabar@}}

\def\ulcorner{\delimiter"4\msafam@70\msafam@70 }
\def\urcorner{\delimiter"5\msafam@71\msafam@71 }
\def\llcorner{\delimiter"4\msafam@78\msafam@78 }
\def\lrcorner{\delimiter"5\msafam@79\msafam@79 }
\def\yen{{\mathhexbox@\msafam@55 }}
\def\checkmark{{\mathhexbox@\msafam@58 }}
\def\circledR{{\mathhexbox@\msafam@72 }}
\def\maltese{{\mathhexbox@\msafam@7A }}

\ifcase\@ptsize
  \font\tenmsb=msbm10
  \font\sevenmsb=msbm7
  \font\fivemsb=msbm5
\or
  \font\tenmsb=msbm10  scaled \magstephalf
  \font\sevenmsb=msbm7 scaled \magstephalf
  \font\fivemsb=msbm5  scaled \magstephalf
\or
  \font\tenmsb=msbm10  scaled \magstep1
  \font\sevenmsb=msbm7 scaled \magstep1
  \font\fivemsb=msbm5  scaled \magstep1
\fi

\newfam\msbfam
\textfont\msbfam=\tenmsb
\scriptfont\msbfam=\sevenmsb
\scriptscriptfont\msbfam=\fivemsb
\edef\msbfam@{\hexnumber@\msbfam}
\def\Bbb#1{{\fam\msbfam\relax#1}}
\def\widehat#1{\setbox\z@\hbox{$\m@th#1$}%
 \ifdim\wd\z@>\tw@ em\mathaccent"0\msbfam@5B{#1}%
 \else\mathaccent"0362{#1}\fi}
\def\widetilde#1{\setbox\z@\hbox{$\m@th#1$}%
 \ifdim\wd\z@>\tw@ em\mathaccent"0\msbfam@5D{#1}%
 \else\mathaccent"0365{#1}\fi}

\ifcase\@ptsize
  \font\teneufm=eufm10
  \font\seveneufm=eufm7
  \font\fiveeufm=eufm5
\or
  \font\teneufm=eufm10   scaled \magstephalf
  \font\seveneufm=eufm7  scaled \magstephalf
  \font\fiveeufm=eufm5   scaled \magstephalf
\or
  \font\teneufm=eufm10   scaled \magstep1
  \font\seveneufm=eufm7  scaled \magstep1
  \font\fiveeufm=eufm5   scaled \magstep1
\fi

\newfam\eufmfam
\textfont\eufmfam=\teneufm
\scriptfont\eufmfam=\seveneufm
\scriptscriptfont\eufmfam=\fiveeufm
\def\frak#1{{\fam\eufmfam\relax#1}}

\csname amssym.def\endcsname

\expandafter\ifx\csname pre amssym.tex at\endcsname\relax \else \endinput\fi
\expandafter\chardef\csname pre amssym.tex at\endcsname=\the\catcode`\@
\catcode`\@=11
\newsymbol\boxdot 1200
\newsymbol\boxplus 1201
\newsymbol\boxtimes 1202
\newsymbol\square 1003
\newsymbol\blacksquare 1004
\newsymbol\centerdot 1205
\newsymbol\lozenge 1006
\newsymbol\blacklozenge 1007
\newsymbol\circlearrowright 1308
\newsymbol\circlearrowleft 1309
\undefine\rightleftharpoons
\newsymbol\rightleftharpoons 130A
\newsymbol\leftrightharpoons 130B
\newsymbol\boxminus 120C
\newsymbol\Vdash 130D
\newsymbol\Vvdash 130E
\newsymbol\vDash 130F
\newsymbol\twoheadrightarrow 1310
\newsymbol\twoheadleftarrow 1311
\newsymbol\leftleftarrows 1312
\newsymbol\rightrightarrows 1313
\newsymbol\upuparrows 1314
\newsymbol\downdownarrows 1315
\newsymbol\upharpoonright 1316
 
\newsymbol\downharpoonright 1317
\newsymbol\upharpoonleft 1318
\newsymbol\downharpoonleft 1319
\newsymbol\rightarrowtail 131A
\newsymbol\leftarrowtail 131B
\newsymbol\leftrightarrows 131C
\newsymbol\rightleftarrows 131D
\newsymbol\Lsh 131E
\newsymbol\Rsh 131F
\newsymbol\rightsquigarrow 1320
\newsymbol\leftrightsquigarrow 1321
\newsymbol\looparrowleft 1322
\newsymbol\looparrowright 1323
\newsymbol\circeq 1324
\newsymbol\succsim 1325
\newsymbol\gtrsim 1326
\newsymbol\gtrapprox 1327
\newsymbol\multimap 1328
\newsymbol\therefore 1329
\newsymbol\because 132A
\newsymbol\doteqdot 132B
 
\newsymbol\triangleq 132C
\newsymbol\precsim 132D
\newsymbol\lesssim 132E
\newsymbol\lessapprox 132F
\newsymbol\eqslantless 1330
\newsymbol\eqslantgtr 1331
\newsymbol\curlyeqprec 1332
\newsymbol\curlyeqsucc 1333
\newsymbol\preccurlyeq 1334
\newsymbol\leqq 1335
\newsymbol\leqslant 1336
\newsymbol\lessgtr 1337
\newsymbol\backprime 1038
\newsymbol\risingdotseq 133A
\newsymbol\fallingdotseq 133B
\newsymbol\succcurlyeq 133C
\newsymbol\geqq 133D
\newsymbol\geqslant 133E
\newsymbol\gtrless 133F
\newsymbol\sqsubset 1340
\newsymbol\sqsupset 1341
\newsymbol\vartriangleright 1342
\newsymbol\vartriangleleft 1343
\newsymbol\trianglerighteq 1344
\newsymbol\trianglelefteq 1345
\newsymbol\bigstar 1046
\newsymbol\between 1347
\newsymbol\blacktriangledown 1048
\newsymbol\blacktriangleright 1349
\newsymbol\blacktriangleleft 134A
\newsymbol\vartriangle 134D
\newsymbol\blacktriangle 104E
\newsymbol\triangledown 104F
\newsymbol\eqcirc 1350
\newsymbol\lesseqgtr 1351
\newsymbol\gtreqless 1352
\newsymbol\lesseqqgtr 1353
\newsymbol\gtreqqless 1354
\newsymbol\Rrightarrow 1356
\newsymbol\Lleftarrow 1357
\newsymbol\veebar 1259
\newsymbol\barwedge 125A
\newsymbol\doublebarwedge 125B
\undefine\angle
\newsymbol\angle 105C
\newsymbol\measuredangle 105D
\newsymbol\sphericalangle 105E
\newsymbol\varpropto 135F
\newsymbol\smallsmile 1360
\newsymbol\smallfrown 1361
\newsymbol\Subset 1362
\newsymbol\Supset 1363
\newsymbol\Cup 1264
 
\newsymbol\Cap 1265
 
\newsymbol\curlywedge 1266
\newsymbol\curlyvee 1267
\newsymbol\leftthreetimes 1268
\newsymbol\rightthreetimes 1269
\newsymbol\subseteqq 136A
\newsymbol\supseteqq 136B
\newsymbol\bumpeq 136C
\newsymbol\Bumpeq 136D
\newsymbol\lll 136E
 
\newsymbol\ggg 136F
 
\newsymbol\circledS 1073
\newsymbol\pitchfork 1374
\newsymbol\dotplus 1275
\newsymbol\backsim 1376
\newsymbol\backsimeq 1377
\newsymbol\complement 107B
\newsymbol\intercal 127C
\newsymbol\circledcirc 127D
\newsymbol\circledast 127E
\newsymbol\circleddash 127F
\newsymbol\lvertneqq 2300
\newsymbol\gvertneqq 2301
\newsymbol\nleq 2302
\newsymbol\ngeq 2303
\newsymbol\nless 2304
\newsymbol\ngtr 2305
\newsymbol\nprec 2306
\newsymbol\nsucc 2307
\newsymbol\lneqq 2308
\newsymbol\gneqq 2309
\newsymbol\nleqslant 230A
\newsymbol\ngeqslant 230B
\newsymbol\lneq 230C
\newsymbol\gneq 230D
\newsymbol\npreceq 230E
\newsymbol\nsucceq 230F
\newsymbol\precnsim 2310
\newsymbol\succnsim 2311
\newsymbol\lnsim 2312
\newsymbol\gnsim 2313
\newsymbol\nleqq 2314
\newsymbol\ngeqq 2315
\newsymbol\precneqq 2316
\newsymbol\succneqq 2317
\newsymbol\precnapprox 2318
\newsymbol\succnapprox 2319
\newsymbol\lnapprox 231A
\newsymbol\gnapprox 231B
\newsymbol\nsim 231C
\newsymbol\ncong 231D
\newsymbol\diagup 231E
\newsymbol\diagdown 231F
\newsymbol\varsubsetneq 2320
\newsymbol\varsupsetneq 2321
\newsymbol\nsubseteqq 2322
\newsymbol\nsupseteqq 2323
\newsymbol\subsetneqq 2324
\newsymbol\supsetneqq 2325
\newsymbol\varsubsetneqq 2326
\newsymbol\varsupsetneqq 2327
\newsymbol\subsetneq 2328
\newsymbol\supsetneq 2329
\newsymbol\nsubseteq 232A
\newsymbol\nsupseteq 232B
\newsymbol\nparallel 232C
\newsymbol\nmid 232D
\newsymbol\nshortmid 232E
\newsymbol\nshortparallel 232F
\newsymbol\nvdash 2330
\newsymbol\nVdash 2331
\newsymbol\nvDash 2332
\newsymbol\nVDash 2333
\newsymbol\ntrianglerighteq 2334
\newsymbol\ntrianglelefteq 2335
\newsymbol\ntriangleleft 2336
\newsymbol\ntriangleright 2337
\newsymbol\nleftarrow 2338
\newsymbol\nrightarrow 2339
\newsymbol\nLeftarrow 233A
\newsymbol\nRightarrow 233B
\newsymbol\nLeftrightarrow 233C
\newsymbol\nleftrightarrow 233D
\newsymbol\divideontimes 223E
\newsymbol\varnothing 203F
\newsymbol\nexists 2040
\newsymbol\Finv 2060
\newsymbol\Game 2061
\newsymbol\mho 2066
\newsymbol\eth 2067
\newsymbol\eqsim 2368
\newsymbol\beth 2069
\newsymbol\gimel 206A
\newsymbol\daleth 206B
\newsymbol\lessdot 236C
\newsymbol\gtrdot 236D
\newsymbol\ltimes 226E
\newsymbol\rtimes 226F
\newsymbol\shortmid 2370
\newsymbol\shortparallel 2371
\newsymbol\smallsetminus 2272
\newsymbol\thicksim 2373
\newsymbol\thickapprox 2374
\newsymbol\approxeq 2375
\newsymbol\succapprox 2376
\newsymbol\precapprox 2377
\newsymbol\curvearrowleft 2378
\newsymbol\curvearrowright 2379
\newsymbol\digamma 207A
\newsymbol\varkappa 207B
\newsymbol\Bbbk 207C
\newsymbol\hslash 207D
\undefine\hbar
\newsymbol\hbar 207E
\newsymbol\backepsilon 237F
\catcode`\@=\csname pre amssym.tex at\endcsname

\def\Box{\hbox{\vrule height1ex\kern-0.4pt
\vbox to 1ex{\hrule width1ex\vfil\hrule width1ex}\kern-0.4pt\vrule height1ex}}
\newcommand{\sqr}[2]{{{\vcenter{\vbox{\hrule height.#2pt
\hbox{\vrule width.#2pt height#1pt \kern#1pt
\vrule width.#2pt}
\hrule height.#2pt}}}}}

\newcommand{\ovl}{\overline}
\newcommand{\til}{\tilde}

\newcommand{\be}{\begin{equation}}

\newcommand{\ee}{\end{equation}}
\newcommand{\al}{\alpha}

\newcommand{\gm}{\gamma}
\newcommand{\dl}{\delta}

\newcommand{\th}{\theta}

\newcommand{\lm}{\lambda}

\newcommand{\rh}{\rho}

\newcommand{\ta}{\tau}

\newcommand{\Ph}{\Phi}
\newcommand{\phv}{\varphi}
\newcommand{\ch}{\chi}

\newcommand{\ps}{\psi}
\newcommand{\Ps}{\Psi}

\newcommand{\raw}{\rightarrow}

\newcommand{\g}{\frak g}

\newcommand{\C}{{\Bbb C}}

\newcommand{\bib}{\bibitem}
\newcommand{\cin}{C^{\infty}}
\renewcommand{\S}{\mbox{$\cal S$}}

\renewcommand{\H}{\mbox{$\cal H$}}

\newcommand{\n}{\parallel}

 \renewcommand{\ll}{\label}

\newcommand{\R}{{\Bbb R}}

\newcommand{\notp}{p \kern-.48em /}
\newcommand{\ci}{\cite}
\newcommand{\bea}{\begin{eqnarray}}
\newcommand{\eea}{\end{eqnarray}}
\newcommand{\ot}{\otimes}
\newcommand{\half}{\mbox{\footnotesize $\frac{1}{2}$}}

\topmargin = - 0.5 cm
\newcommand{\la}{\langle}
\newcommand{\ra}{\rangle}
\newcommand{\CA}{{\cal A}}
\newcommand{\CG}{{\cal G}}
\newcommand{\CH}{{\cal H}}
\newcommand{\CN}{{\cal N}}
\newcommand{\Sp}{S^{\rm phys}}
\newcommand{\Hp}{{\cal H}^{\rm phys}}
\newcommand{\T}{\Bbb T}
\renewcommand{\t}{{\frak t}}
\newcommand{\Z}{\Bbb Z}
\newcommand{\mip}{\langle \, | \, \rangle_{\rm phys}}
\newcommand{\mipo}{\langle \, | \, \rangle_0}
\newcommand{\bg}{{\bf g}}
\newcommand{\Exp}{{\rm Exp}}

\newcommand{\rep}{representation}
\begin{document}
\setlength{\baselineskip}{1.5\baselineskip}
\thispagestyle{empty}
\title{Constrained quantization and $\th$-angles}
\author{N.\ P.\ Landsman\thanks{ E.P.S.R.C.\ Advanced Research Fellow}\mbox{ }
and K.K. Wren
\\
  Department of Applied Mathematics and Theoretical Physics\\ University of
Cambridge\\ Silver Street, Cambridge CB3 9EW, U.K. }
\date{\today}
\maketitle
\begin{abstract}
 We apply a new and mathematically rigorous method for the
quantization of constrained systems to two-dimensional gauge
theories. In this method, which quantizes Marsden-Weinstein symplectic
reduction, the inner product on the physical state space is expressed
through a certain integral over the gauge group.  The present paper,
the first of a series, specializes to the Minkowski theory defined on
a cylinder.  The integral in question is then constructed in terms of
the Wiener measure on a loop group. It is shown how $\th$-angles
emerge in the new method, and the abstract theory is illustrated in
detail in an example.
\end{abstract}
\newpage
\section{Introduction}
\subsection{Classical reduction}
In Dirac's theory of constrained dynamical systems \ci{Dir,GNH,BSF}
the so-called {\em reduced phase space} is generically obtained by a two-step
reduction procedure. In  summary, the two steps of the reduction of a
classical constrained
system are 
\begin{enumerate}
\item 
Imposing the constraints $\Ph_i=0$; this restricts the phase space of
the unconstrained system $S$ to 
the  {\em constraint hypersurface} $C$.
\item
Quotienting by {\em the null foliation} $\CN_0$ of the induced
 symplectic form on the 
 constraint hypersurface $C$.
\end{enumerate}
The reduced phase space is then $S^0=C/\CN_0$. This often, but not
always, coincides with the  
phase space $\Sp$ of physical degrees of freedom.

Roughly speaking, the second step undoes the underdetermination of the
equations of motion on $C$; in gauge theories with connected gauge
group physically equivalent points are identified by this step.
Indeed, in a gauge theory (formulated in the temporal gauge for
simplicity) the constraints are given by Gauss' law, and quotienting
by the null foliation amounts to collapsing each orbit of the identity
component $\CG_0$ of the (time-independent) gauge group $\CG$ to a
point; one has $S^0=C/\CG_0$.  If $\CG$ is not connected, one needs to
include a further step in order to arrive at $\Sp$, viz.\ quotienting
$C/\CG_0$ by the discrete group $\pi_0(\CG)=\CG/\CG_0$. Thus
$\Sp=S^0/\pi_0(\CG)=C/\CG$.

More generally, consider the case that a Lie group $\CG$ acts
canonically on $S$ (that is, the action preserves the Poisson
bracket). In the absence of certain topological obstructions
\ci{GS,AM} this action is then generated by functions $\Ph_i$ (chosen
relative to a basis $\{T_i\}$ of the Lie algebra $\bg$ of $\CG$) on
$S$, whose Poisson brackets reproduce the Lie bracket in $\bg$; i.e.,
$\{\Ph_i,\Ph_j\}=C_{ij}^k\Ph_k$.  Each $\Ph_i$ plays the role of a
charge, and it often happens that constraints are given by $\Ph_i=0$
for such charges. This setting, indeed, applies in the case of a gauge
group \ci{Arms2,BSF}; also see section \ref{mwr} below.  We will refer
to this situation as {\em the group case}; the associated reduction of
$S$ is known as {\em Marsden-Weinstein reduction} \ci{GS,AM,BSF}.
\subsection{Dirac's quantum reduction} In trying to find a quantum
analogue of the classical reduction procedure, Dirac \ci{Dir} saw that
only one of the two classical steps needs to be `quantized'. Let us
restrict ourselves to the case where all constraints are first-class
(this means that all Poisson brackets $\{\Ph_i,\Ph_j\}$ vanish on
$C$); this special case is the heart of the matter, and includes gauge
theories.  Assume that, through some construction, a Hilbert space
$\H$ is given as the quantization of the (unconstrained) classical
phase space $S$. Along with $\H$, which serves as the quantum state
space of the unconstrained system, suppose the classical constraints
have been quantized into operators $\hat{\Ph}_i$ on $\H$.

 Dirac, then, proposed that the quantization of $S^0$ be given by \be
\H^0_D:=\{|\ps\ra\in\H|\:\:\: \hat{\Ph}_i|\ps\ra=0\:\forall i\}; \ee
that is, $\H^0_D$ is the subspace of $\H$ which is annihilated by the
quantum constraints.  It inherits the inner product from $\H$, so that
it becomes a Hilbert space in its own right, in which physical
amplitudes may be computed.  Consistency of this proposal entails that
each commutator $[\hat{\Ph}_i,\hat{\Ph}_j]$ must annihilate $\H_D$
(i.e., the quantum theory is anomaly-free).

In the group case it suffices that the $\hat{\Ph}_i$ form a
representation of the Lie algebra $\bg$. This Lie algebra
representation usually corresponds to a unitary representation $U$ of
$\CG$, in which case the space $\H^0_D$ may alternatively be
characterized as \be \H^0_D=\{|\ps\ra\in\H|\:\:\: U(g)|\ps\ra=|\ps\ra
\:\forall g\in\CG_0\}. \ll{dcg} \ee This evidently leaves open the
question how, in case $\CG$ is disconnected, the Hilbert space $\Hp$
is to be defined.
 
As we see, there is no analogue of the quotienting step of classical
reduction, which would, in a way, render quantum reduction a simpler
procedure than its classical counterpart.  The reader will now remark
that the quantum BRST procedure, at least in its operator version,
does mimic its classical counterpart in being a two-step procedure as
well. This is not the place to point out at what cost this is achieved
\ci{LL}; the relevant point is that the first step in quantum BRST
leads to problems entirely similar to the ones encountered in the
Dirac approach.

Dirac's proposal has particularly dominated the literature on
canonical quantum gravity and quantum cosmology, where the so-called
Hamiltonian constraint implies the controversial Wheeler-DeWitt
equation. The difficulties this equation leads to are by now widely
known and acknowledged \ci{Ish,Ash}, although it is not always
appreciated that most of these are merely a special instance of
general problems with the Dirac (and operator-BRST) approach. The main
difficulties are:
\begin{itemize}
\item
It is very rare that all quantum constraints have 0 in their discrete
spectrum, with joint eigenspace. In other words, the equations
$\hat{\Ph}_i|\ps\ra=0$ often have no solution in $\H$.  This situation
usually occurs when the group generated by the constraints is not
compact.
\item
If one seeks solutions outside $\H$, one has to construct an inner
product on the space of solutions afresh. While this is possible in
certain cases, there is no good prescription as to which (generalized)
solutions to include.
\end{itemize}
In quantum cosmology the last problem lies behind the discussion what
the `wave function of
the universe' should be \ci{WDW}.
\subsection{A new method of constrained quantization}\label{rfl}
In view of these difficulties, and also for purely mathematical
reasons, alternatives to Dirac's quantization procedure (or its BRST
version) have been sought.  We shall here make use of one such
alternative \ci{L2,L3}\footnote{Related methods will be mentioned at
the end of this subsection.}, whose essential idea is to quantize the
second rather than the first step of classical reduction.  This new
approach turns out to work even when the Dirac (or BRST) method breaks
down, reducing to it in those cases where it happens to apply.  Also,
one has a clean definition and construction of (weak) quantum
observables (see below).
 
 In its simplest version, this idea is implemented by manipulating the
inner product $\la\, |\,\ra$ on $\H$ (which by definition is positive
definite) into a sesquilinear form $\la\, |\,\ra_{\rm phys}$ which is
positive semidefinite.  The construction of this form is dictated by
the constraints.  The form $\la\, |\,\ra_{\rm phys}$ will have a
nonempty null space \be {\cal N}=\{|\ps\ra\in\H|\:\:\:\la\ps
|\ps\ra_{\rm phys}=0\}; \ee the physical state space is then given by
\be \H^{\rm phys}=\H/\CN.  \ee The inner product $\la\, |\, \ra^{\rm
phys}$ on $\H^{\rm phys}$ is the one inherited from $\la\, |\,\ra_{\rm
phys}$; it is positive definite by construction.  If $V:\H\raw \H^{\rm
phys}$ is the canonical projection, one therefore has \be
\la\ps|\phv\ra_{\rm phys}=\la V\ps |V\phv \ra^{\rm phys}. \ll{VIP} \ee
The Hilbert space $\H^{\rm phys}$ is the quantization of $\Sp$. There
is no need to pass through an intermediate space $\H^0$ (quantizing
$S^0$), although it often provides insight to do so.

The set of bounded {\em weak quantum observables} consists of those
bounded operators $B$ on $\H$ which are self-adjoint with respect to
the manipulated inner product, i.e., which satisfy \be \la
\ps|B|\phv\ra_{\rm phys}= \ovl{\la \phv|B|\ps\ra}_{\rm phys}\ll{weak}
\ee for all $\ps,\phv\in\H$ (here the bar stands for complex
conjugation).  Without the subscript `phys' this would, of course, be
the condition that $B$ be Hermitian.  A weak quantum observable $B$
maps $\CN$ into itself, so that its `induced' action on the quotient
$\H^{\rm phys}$ specifies a well-defined physical observable $B^{\rm
phys}$. By definition, one has \be B^{\rm phys} V|\ps\ra=VB|\ps\ra,
\ll{Aphys} \ee and this property completely specifies $B^{\rm phys}$
as an operator on $\Hp$.

In practice $\la\, |\,\ra_{\rm phys}$ is often only well-defined on a
certain dense subspace ${\cal D}\subset\H$; this happens precisely
when the Dirac procedure breaks down.  In that case the above
construction of $\H^{\rm phys}$ undergoes only minor modifications:
the null space $\CN$ is now defined as a subspace of ${\cal D}$, and
the quotient ${\cal D}/\CN$ has to be completed in the inner product
$\la\, |\,\ra^{\rm phys}$ to obtain $\H^{\rm phys}$.  With this
refinement, the key mathematical problems in the Dirac or BRST
approaches are avoided.  All this even works if all constraints are
second-class; in fact, the classification of the constraints into
first- and second-class ones is unnecessary in our procedure.

In this more general case, a weak quantum observable $B$ is a possibly
unbounded operator whose domain contains ${\cal D}$, and which leaves
${\cal D}$ stable. As in the previous paragraph, when $B$ is a weak
quantum observable the induced operator $B^{\rm phys}$ on $\Hp$ is
well-defined, and represents a physical quantum observable.

Let us return to the group case, supposing that the quantum
constraints generate a unitary representation $U(\CG)$ on $\H$. The
construction of the manipulated inner product for this situation is
explained in detail in \ci{L2,L3}, with the following result. For the
moment we assume that $\CG$ is connected. If $\CG$ is compact, one has
${\cal D}=\H$, and \be \la\ps| \phv\ra_{\rm phys}=\int_{\CG_0} dg\,
\la \ps|U(g)|\phv\ra, \ll{ip} \ee where $dg$ is the Haar measure; for
later reference we have written $\CG_0$ for $\CG$ to reflect its
connectedness.  This equals $\la \ps |P_{\rm id}|\phv\ra$, where
$P_{\rm id}$ is the projector on the subspace $\H_{\rm id}$ of $\H$
which transforms trivially under $U(\CG)$. Hence $\CN=\H^{\perp}_{\rm
id}$, so that $\H^{\rm phys}\simeq \H_{\rm id}$. This coincides with
$\H_D$ of the Dirac method; cf.\ (\ref{dcg}).  A crucial property of
the manipulated inner product, which is immediate from the above, is
\be \la \psi|U(h)|\phv\ra_{\rm phys}=\la \ps|\phv\ra_{\rm phys}
\ll{cp} \ee for all $h\in\CG$ and all $\psi,\phv\in\H$.  According to
(\ref{weak}) and (\ref{cp}), each $U(h)$ is a weak quantum observable,
which by (\ref{cp}) and (\ref{Aphys}) is represented by the unit
operator on $\Hp$. This suffices to prove that $\CG_0$ acts trivially
in the physical space $\Hp$.

If $\CG$ is merely locally compact, but not compact, (and here assumed
unimodular for simplicity, so that the left- and right- Haar measure
coincide) the expression (\ref{ip}), and its consequence (\ref{cp})
still follows, but is only defined on a suitable dense domain ${\cal
D}\subset\H$. The projection $P_{\rm id}$ and the space $\H_{\rm id}$
no longer exist (so that the Dirac approach would break
down). However, one can successfully proceed as indicated earlier.

The case were $\CG$ is not even locally compact, e.g.\ if $\CG$ is a
gauge group, will be faced in the present paper (also cf.\
\ci{LWied,WiedL}).  It turns out that one can still make good
mathematical sense of an expression of the above kind, despite the
non-existence of Haar measures on infinite-dimensional groups.
 
The idea of group-averaging in the context of constrained quantization
goes back, at least, to Teitelboim \ci{Tei1,Tei2}; it is, of course,
common practice in lattice gauge theory.  The constrained quantization
procedure proposed in \ci{Ash} also involves expressions of the type
(\ref{ip}).
\subsection{Discrete reduction and $\th$-angles\ll{discrete}}
As already remarked, the case where the gauge group $\CG$ is
disconnected is exceptional  
in that the reduced phase space $S^0=C/\CG_0$ (although symplectic)
does not coincide with the 
physical phase space $\Sp=S^0/\pi_0(\CG)$. The passage from $S$ to
$S^0$ rather than $\Sp$ can be 
mimicked in quantum theory by restricting the integral in (\ref{ip})
to $\CG_0$; this leads to a 
Hilbert space $\H^0$, which is the quantization of $S^0$. The passage
from $\H^0$ to $\Hp$ would then 
involve a step `quantizing' the passage from $S^0$ to $\Sp$. One
could, of course, postulate 
(\ref{ip}) also for disconnected $\CG$, but this would overlook an
important option one has available at this point.

We isolate the issue at stake by looking at the classical reduction of
an arbitrary symplectic manifold $S^0$ by a discrete group $D$
\ci{Mar}.  Unlike in general reduction, there is only one step, namely
the passage to the quotient $S^0/D$. When $D$ acts freely on $S^0$
this is a manifold, but if it doesn't $S^0/D$ is typically an orbifold
(locally a manifold) \ci{Sat}.  More importantly, $S^0/D$ is
symplectic (away from its possible singular points), quite unlike a
quotient by a connected Lie group, which would not be symplectic,
being merely a Poisson space \ci{Mar}.

Accordingly, there are no constraints, and $C=S^0$. The absence of
constraints leads to a certain freedom in the definition of the
manipulated inner product $\mip$. Namely, having quantized $S^0$ by a
Hilbert space $\H^0$, carrying a unitary representation $U^0$ of $D$,
as before, we could pick an arbitrary one-dimensional unitary
representation $U_{\th}$ of $D$ (where $\th$ stands for a collection
of parameters labelling such representations), and define the
manipulated inner product on $\H$ (or on a suitable dense subspace) by
\be \la\ps|\phv\ra^{\th}_{\rm phys}=\sum_{g\in D} U_{\th}(g)
\la\ps|U^0(g)|\phv\ra. \ll{ip3} \ee The freedom to include $U_{\th}$
is due to the fact that in discrete classical reduction the ``0'' in
$\Ph_i=0$, which would force $U_{\th}$ to be trivial, is absent.

In case that $D$ is nonabelian, the above limitation to
one-dimensional representations $U_{\th}$ yields $\th$-parameters only
if $D/[D,D]$ is nontrivial.  In the framework of \ci{L2,L3} it is, in
fact, entirely possible to work with general unitary representations
of $D$ (also cf.\ \ci{Giu}).  If $U_{\th}$ is defined on some Hilbert
space $\H_{\th}$, one defines the manipulated inner product $\mip$ on
$\H\otimes\H_{\th}$ rather than on $\H$ (or, if need be, on ${\cal
D}\otimes\H_{\th}$) by the obvious generalization of (\ref{ip3}),
viz. by sesquilinear extension of \be \la\ps\otimes v|\phv\otimes
w\ra_{\rm phys}=\sum_{g\in D} \la v|U_{\th}(g)|w\ra
\la\ps|U(g)|\phv\ra, \ll{ip4} \ee where $\ps,\phv\in\H$ and $v,w\in
\H_{\th}$.  The construction of $\Hp$ then proceeds as before; the
null space $\CN$ is now a subspace of $\H\otimes\H_{\th}$ (or ${\cal
D}\otimes\H_{\th}$). This is, for example, relevant for braid group
statistics.

Returning to the general group case, with $\CG$ disconnected, we can
proceed at one stroke, avoiding the intermediate space $\H^0$, by
replacing (\ref{ip}) by \be \la\ps| \phv\ra^{\th}_{\rm
phys}=\int_{\CG} dg\, \til{U}_{\th}(g)\la \ps|U(g)|\phv\ra, \ll{ip2}
\ee where $\til{U}_{\th}(g)=U_{\th}\circ\ta_{\CG\raw \CG/\CG_0}$ is
defined through a one-dimensional unitary representation $U_{\th}$ of
$\CG/\CG_0$; here $\ta_{\CG\raw \CG/\CG_0}$ is the canonical
projection from $\CG$ to $\CG/\CG_0$.  The inner product on the
physical Hilbert space of states $\Hp$ will then depend on $\th$,
although this $\th$-dependence is usually undone by a unitary
transformation which puts it into the physical observables.

This procedure is equivalent to the following one, which in effect
breaks the process up into two separate steps. Hence one mimicks the
classical reduction process, and obtains an algorithm that in practice
is often easier to implement.  One first constructs $\H^0$ using
(\ref{ip}); this only uses the restriction of $U(\CG)$ to
$\CG_0$. Since $\CG_0$ acts trivially on $\H^0$ by construction, the
full representation $U(\CG)$ on $\H$ quotients to a representation
$U^0(\CG/\CG_0)$ on $\H^0$.  One then puts the manipulated inner
product (\ref{ip3}) (with $D =\pi_0(\CG)$) on $\H^0$, and proceeds to
construct the physical state space $\Hp$ as before.

In other treatments of `$\th$' phase factors \ci{Shu,Ish2,HMS,Giu},
restricted to the case of a multiply connected configuration space
$Q$, one encounters the fundamental group $\pi_1(Q)$. To relate this
to $\pi_0(\CG)$, note that a multiply connected space $Q$ may be
written as $Q=\ovl{Q}/D$, with $D=\pi_1(Q)$, and $\ovl{Q}$ is the
universal covering space of $Q$. By construction, $\pi_1(\ovl{Q})=e$,
and if $Q$ is connected one has the equality $\pi_1(Q)=\pi_0(D)$.
Hence we choose $S^0=T^*\ovl{Q}$, on which $D$ acts by pull-back; the
reduced space is $T^*\ovl{Q}/D\simeq T^*Q$ (see below). In the
opposite direction, we could start from some connected and simply
connected space $X$ (the configuration space of gauge fields, which is
affine, being a case in point) and reduce $S^0=T^*X$ by the action of
a discrete group $D$ on $X$ (pulled back to $T^*X$).  Since $(T^*X)/D
\simeq T^*(X/D)$ for discrete $D$, the above-mentioned approaches
would look at this as the problem of quantizing the multiply connected
space $Q=X/D$. By the same argument we have $\pi_0(D)=\pi_1(X/D)$.
 
In all other respects our way of introducing $\th$-angles is
profoundly different from others, and one goal of this paper is to
explicitly illustrate how these angles emerge in a mathematically
rigorous constrained quantization method. In the context of gauge
theories the two methods of explaining such angles that are best known
to field theorists (cf.\ \ci{Jac,Wei2} for reviews) are, so to speak,
`orthogonal' to ours. Firstly, in the (Euclidean) path-integral
method, where the $\th$-parameter enters through instantons, one does
not integrate over the gauge group: it is precisely the goal of the
Faddeev-Popov gauge fixing procedure to {\em avoid} such an
integration. In our approach, on the other hand, all effects come from
this integration.  Secondly, in the Hamiltonian approach one
postulates that physical states transform like
$U(g)|\ps\ra=\til{U}^{-1}_{\th}(g)|\ps\ra$; this generalization of
Dirac's condition $U(g)|\ps\ra=|\ps\ra$ (cf.\ (\ref{dcg})) is avoided
here, for the same reasons that Dirac's original condition is
bypassed.  \subsection{Gauge theory on a circle} We will illustrate
the new technique for gauge theories on a circle $\T\simeq \R/2\pi\Z$.
These resemble topological field theories in that the physical phase
space is finite-dimensional; see \ci{Man} for the abelian case and
\ci{Raj,Raj2,Raj3} for the compact nonabelian case.  In fact, the
physical configuration space of a pure Yang-Mills theory on a circle
is $G/{\rm ad}(G)$ (that is, the space of orbits of the adjoint action
of $G$ on itself); the derivation of this result by Marsden-Weinstein
reduction will be reviewed below.  For connected $G$ this space is
diffeomeorphic to $T/W$, where $T$ is a maximal torus in $G$ and $W$
is the associated Weyl group; see \ci{Raj2} for a rigorous derivation
in the present context, and also cf.\ Theorem 4.44 in \ci{knapp} for
the isomorphism $G/{\rm ad}(G)\simeq T/W$.  This space is singular
(but note that $T/W$ is not an orbifold in the sense of \ci{Sat}), and
some care is needed in the definition of the cotangent bundle
$T^*(G/{\rm ad}(G))$; with the correct definition this is the physical
phase space $\Sp$.

To quantize according to our method, we have to
face the   full complexity of the problem of defining the integral in (\ref{ip})
or (\ref{ip2}).  It turns out that the correct choice of the gauge group $\CG$ is to include all
continuous loops in $G$ with finite energy; in analogy with the situation on flat space  \ci{CM}
we might call $\CG$ the  Cameron-Martin loop group. This choice,
however, leads to two (apparent) difficulties.

Firstly, being infinite-dimensional, the gauge group $\CG$ has no Haar
measure. It turns out that, heuristically speaking, the would-be
``Haar measure'' $dg$ on $\CG$ combines with a Gaussian factor in the
matrix element of $U(g)$ to form a well-defined measure. This
combination closely resembles the way the non-existent Lebesgue
measure on the space of paths in $\R^3$ combines with the exponential
of the kinetic term in the Euclidean action to form the Wiener measure
appearing in the Feynman-Kac formula (cf., e.g., \ci{Sim}).  Hence one
obtains essentially the Wiener measure $\mu_W$ (conditioned on
loops). The Wiener measure on a loop group has appeared in the
literature before in various different contexts; see \ci{AHK,FRKL,MM}.

The second difficulty is, then, that $\CG$ has measure zero w.r.t.\
$\mu_W$. While this may appear paradoxical to physicists, it is simply
the well-known phenomenon that paths with finite energy are too
regular to be supported by the Wiener measure.  Instead of integrating
over $\CG$ in (\ref{ip}) and (\ref{ip2}), we therefore integrate over
the closure $\ovl{\CG}$ of $\CG$ in a natural norm. This closure is
simply the space of all continuous loops. The representation $U(\CG)$
cannot be extended to $\ovl{\CG}$, but such an extension is not needed
to define the manipulated inner product.
 
Using the Wiener measure on $\ovl{\CG}$, the manipulated inner product
can be computed explicitly, and the structure of the physical Hilbert
space $\Hp$ and the action of physical observables may be derived.

In the present paper we perform this computation when the structure
group $G$ is $U(1)$. In that case the gauge group $\CG$ of maps from
$\T$ to $U(1)\simeq \T$ is disconnected, with
$\pi_0(\CG)=\pi_1(G)=\Bbb Z$ (the gauge transformations are labelled
by their winding number). As far as $\pi_0$ is concerned this mimicks
the case where space is $S^3$ and $G=SU(2)$ (for here
$\pi_0(\CG)=\pi_3(G)=\Bbb Z$).

 The construction of $\Hp$ for compact semi-simple $G$ requires
special techniques and involves fascinating mathematics, which is
beyond the scope of the present paper; see \ci{KKW} for a detailed
treatment.  As expected, the physical Hilbert space comes out to be
$\Hp=L^2(G/{\rm ad}(G))$ (cf.\ \ci{Raj,dimock} for other approaches to
the quantization of the Minkowski version of this model, and
\ci{Wit1,Wit2,Sen} for the Euclidean version), but the point of the
derivation lies not so much in the result as in the method.
\subsection{Acknowledgement}
The authors are profoundly indebted to Brian C. Hall for patiently
clarifying L. Gross' approach to the Wiener measure, allowing them to
correct a highly misleading statement in the first draft of this
paper.  \section{Classical reduction} In this section we perform the
Marsden-Weinstein reduction of the unphysical phase space $S$ of
Yang-Mills theory on a circle $\T$ to the physical phase space
$\Sp$. We assume that the structure group $G$ is a connected compact
Lie group, whose Lie algebra is denoted by $\g$.  Without loss of
generality we take the principal $G$-bundle over $\T$, which defines
the classcial setting, to be trivial, i.e., $P=\T\times G$. We
formulate the theory in the temporal gauge $A_0=0$ from the start;
this partial gauge fixing is entirely innocent, and allows us to
regard the gauge group $\CG$ as consisting of maps from $\T$ to
$G$. The configuration space $\CA$ consists of certain functions from
$\T$ to $\g$. The action $g:A\raw gA$ of $g\in\CG$ on $A\in\CA$ is
given by \be gA(\al)={\rm Ad}(g(\al))A(\al)-dg(\al)\, g^{-1}(\al),
\ll{GonA} \ee where ${\rm Ad}(x)A= xAx^{-1}$ stands for the adjoint
action of $G$ on $\g$, and $\al\in\T$.

For the basic mathematical structure of gauge theories we refer to
\ci{AMP}; functional-analytic aspects are covered by
\ci{MV,FU,KS}. Refs.\ \ci{AMP} and \ci{FU} also contain most other
mathematical prerequisites for this chapter.
\subsection{Choice of the gauge group\label{CGU}}
It is necessary to be quite precise about the nature of the spaces
$\CG$ and $\CA$.  The gauge group $\CG$, whose choice dictates that of
$\CA$, should not be too large, in that a space containing
discontinuous gauge transformations would not reflect the topology of
the bundle $P$.  On the other hand, it should not be too small, since
gauge transformations and connections that are too smooth cannot be
used as the basis of a quantum theory.  We will choose $\CG$ to be the
largest subspace of the group of all continuous loops for which both
classical reduction can be successfully carried out, {\em and} the
unitary representation $U(\CG)$ lying at the heart of the construction
of the quantum theory is well defined.

To define $\CG$ we recall that a compact Lie group has a Riemannian
structure obtained by choosing an Ad-invariant Euclidean inner product
$(\, ,\,)_{\g}$ on $\g\simeq\R^n$, and translating this from $\g=T_eG$
to the tangent space of other points by the group action.  Hence for a
curve $\gm:[0,2\pi]\raw G$ (and in particular for a loop $g$) we can
define the function $|\dot{g}|:[0,2\pi]\raw \R$. The space $H_1(\T,G)$
by definition consists of those $g\in C(\T,G)$ whose (weak) derivative
$\dot{g}$ is square-integrable in that $|\dot{g}|\in
L^2([0,2\pi],\R)$. (Here and in what follows, $L^2([0,2\pi],\dots)$ is
defined w.r.t\ the Lebesgue measure $d\al$, as distinct from
$L^2(\T,\ldots)$ which is defined w.r.t.\ the normalized measure
$d\al/2\pi$.)  In particular, the Riemannian length of $g\in
H_1(\T,G)$ exists. It can also be shown that such a $g$ is absolutely
continuous, and that $\dot{g}$ exists almost everywhere; see
\ci{Kli}. Physically, one could say that $H_1(\T,G)$ consists of all
continuous loops with finite energy.

An alternative characterization of $H_1(\T,G)$ is to take the defining
representation $U_d(G)$
on $\H_d$; the space $M_d$ of matrices on $\H_d$ is a normed space, so
that one can define
the Hilbert space $H_1(\T,M_n)$ as the completion of $\cin(\T,M_n)$  in the
$p=1$ Sobolev norm. Then  $H_1(\T,G)$ is the subspace of $H_1(\T,M_n)$
consisting of those functions
which take values in $U_d(G)$. This endows  $H_1(\T,G)$ with the structure of
  a Hilbert manifold (cf.\ \ci{AMP,FU}). The continuous inclusion
$H_1(\T,G)\subset C(\T,G)$ is then
a consequence of the  Sobolev embedding theorem (cf.\ \ci{FU}), from
which it also follows
that $H_1(\T,G)$ is not contained in any $C^p(\T,G)$ for $p>0$.
 
{\em The gauge group is the Hilbert Lie group \be \CG=H_1(\T,G) \ee
with Lie algebra\footnote{Generically Hilbert spaces are over the
complex numbers, unless a real vector space is explicitly indicated,
as in $L^2([0,2\pi], \g)$ or $ H_1(\T,\g)$, which are real Hilbert
spaces.}  \be \bg= H_1(\T,\g).  \ee The group operations in $\CG$ are
pointwise multiplication and inverse; these are smooth with respect to
the Hilbert manifold structure of $\CG$. }
 
For the last point see \ci[App.\ A]{FU}.  Here $H_1(\T,\g)$ is defined
analogously to $H_1(\T,G)$; it is a Hilbert space under the $p=1$
Sobolev inner product \be \la f,g\ra_1=\int_{\T}d\al\, \left(
(f(\al),\ovl{g(\al)})_{\g}+
(\dot{f}(\al),\ovl{\dot{g}(\al)})_{\g}\right).  \ee One has the
inclusion $H_1(\T,\g)\subset C(\T,\g)$, and the pointwise exponential
map on $\bg$ is continuous \cite{FU}.

The connectivity properties of $\CG$ are determined by the following result.

{\em With the gauge group $\CG$ defined as in \ref{CGU}, and the
structure group $G$ equipped with its usual topology as a Lie group,
one has} \be \pi_0(\CG)=\pi_1(G) .\ll{pio} \ee
 
To put this in perspective, note that one usually considers the loop
  group $LG= C(\T,G)$, equipped with the topology of uniform
  convergence (with respect to the metric topology of $G$ inherited
  from the Riemannian structure, or from $G\simeq U_d(G)$ as above).
  This topology coincides with the compact-open topology, so that one
  has \be \pi_0(LG)=\pi_1(G) \ll{pi0lg} \ee by definition of $\pi_1$.

For example, if $G=U(1)$ it follows that $\pi_0(LG)=\Bbb Z$; the
members of a given component $LG_n$, $n\in \Bbb Z$, are labelled by
the winding number of the loop.  More generally, $\pi_1(G)$ is
isomorphic to a discrete subgroup $D$ of the center of the universal
covering group $\ovl{G}$ of $G$ (i.e. $G=\ovl{G}/D$).  Under this
isomorphism an element $[\dl]\in \pi_1(G)$ is the equivalence class of
loops in $G$ which are homotopic to the projection (from $\ovl{G}$ to
$G$) of a path from $e$ to $\dl$ in $\ovl{G}$.

Using (\ref{pi0lg}), we label the
components $LG_{\dl}$ of $LG$ by $\dl\in D$.    
 Since the inclusion $\CG\subset LG$ is continuous with respect
to the manifold topology on $\CG$, (\ref{pio})
 will follow if
each intersection $\CG_{\dl}=\CG\cap LG_{\dl}$ is connected in the topology of
$\CG$; we write $\CG_0$ for $\CG_e$. To prove this,  
by the reasoning in the previous paragraph it suffices to show that
any two $H_1$-paths
in $\ovl{G}$ between $e$ and $\dl$ are homotopy-equivalent in $H_1$, which
is obvious. Hence (\ref{pio}) follows.

An explicit description of a component $\CG_{\dl}$ of $\CG$ is as
follows.  Using the fact that the exponential map $\ovl{\Exp}:\g\raw
\ovl{G}$ is surjective for compact connected Lie groups \ci{knapp}, we
can find a $X_{\dl}\in\g$ for which $\ovl{\Exp}(X_{\dl})=\dl$. If
$[x]_D$ denotes the equivalence class in $G=\ovl{G}/D$ of
$x\in\ovl{G}$, we have \be \CG_{\dl}=\CG_0 g_{\dl}, \ll{d1} \ee where
$g_{\dl}(\al)=[\ovl{\Exp}(X_{\dl}\al/2\pi)]_D$. In other words, any
element $g_{(\dl)}$ of $\CG_{\dl}$ is of the form \be
g_{(\dl)}(\al)=[\ovl{\Exp}(\lm(\al)+X_{\dl}\al/2\pi)]_D, \ll{d2} \ee
where $\lm\in \bg$; in particular, $\lm(2\pi)=\lm(0)$.

For example, if $G=\T$ one has $D=\pi_0(G)=2\pi\Z$; one usually labels
elements of $\T$ by $\al\in[0,2\pi)$. The Lie algebra $\t$ of $\T$ as
well as of its covering group $\ovl{\T}=\R$ is identified with $\R$;
then $\Exp:\t\raw \T$ is given by $\Exp(X)=\exp(iX)$, whereas
$\ovl{\Exp}:\t\raw\R$ is the identity map.  Hence $\dl\in\R$ is of the
form $2\pi n$; we then have, with slight abuse of notation, \be
g_n(\al)=e^{in\al}. \ll{gnal} \ee
 
Finally, we determine the appropriate space of connections $\CA$; our
choice is the same as the one in \ci{Raj2}.  If $g\in H_1(\T,G)$ then
$dg\, g^{-1}\in H_0(\T,\g)=L^2([0,2\pi],\g)$. Hence we choose
$\CA=L^2([0,2\pi],\g)$ (a real Hilbert space). It can be shown that
the action of $\CG$ on $\CA$ is smooth \ci[App.\ A]{FU}. Since $\CA$
is a Hilbert space, the cotangent bundle is
$T^*\CA=L^2([0,2\pi],\g^*)\times L^2([0,2\pi],\g)$.  We write elements
of $S$ as pairs $(E,A)$, where $E$ and $A$ take values in $\g^*$ and
$\g$, respectively.  The $\CG$-action on $\CA$ (\ref{GonA}) lifts to a
smooth $\CG$-action on $S$ given by \be g:(E,A)\raw ({\rm Co}(g)E,
{\rm Ad}(g)A-dg\, g^{-1}), \ll{groupaction} \ee where we have omitted
the argument $\al$, and ${\rm Co}$ stands for the co-adjoint action of
$G$ on $\g^*$.  Note that $dg\, g^{-1}$ is not, in general, an element
of $\bg$.  The infinitesimal transformation generated by $\lm\in\bg$
is \be \lm: (E,A)\raw (E+{\rm Co}(\lm) E, A-D_A\lm), \ll{infgt} \ee
where ${\rm Co}(\lm)$ stands for $\lm$ taken in the co-adjoint
representation, and $D_A\lm=d\lm+[A,\lm]=d\lm-{\rm Ad}(\lm)A$.  We may
identify $\g$ with its dual $\g^*$ through the choice of an inner
product on $\g$; then ${\rm Co}(\lm) E$ is replaced by ${\rm
Ad}(\lm)E=[\lm,E]$.  \subsection{Marsden-Weinstein reduction\ll{mwr}}
The procedure of Marsden-Weinstein reduction is well-defined also for
infinite-di\-men\-sional (strongly) symplectic manifolds \ci{AM}; see
in particular \ci{Arms2,BSF} for Marsden-Weinstein reduction in the
context of gauge theories.  We here take $S=T^*\CA$, and reduce with
respect of the group action (\ref{groupaction}). The Poisson bracket
on $\cin(S)$ is given by \be \{F,G\}=\int_{\T} d\al \, \left(
\frac{\dl F}{\dl E_a(\al)}\frac{\dl G}{\dl A^a(\al)}- \frac{\dl F}{\dl
A^a(\al)}\frac{\dl G}{\dl E_a(\al)}\right) , \ll{pb} \ee where
$A=A^aT_a$ and $E=E_a \th^a$ in terms of a basis $\{T_a\}$ of $\g$ and
its dual basis $\{\th^a\}$ of $\g^*$.  For the linear functionals
$F(A)=A(f)=\la f|A\ra$ and $G(E)=E(g)=\la g|E\ra$ on $S$, where
$f,g\in \CA$ are smearing functions, (\ref{pb}) yields \be
\{A(f),E(g)\}=-\la f|g\ra. \ll{pbAE} \ee In particular,
$\{A(1),E(1)\}=2\pi$.
  
It is clear that the action (\ref{groupaction}) preserves this Poisson
 bracket, so that it is canonical.  A {\em momentum map} $J$ is a
 function from $S$ to the dual Lie algebra $\bg^*$, which by
 definition satisfies \be \{J_{\lm},f\}= \dl_{\lm}f; \ee we write
 $J_{\lm}$ for $\la J,\lm\ra$, where $\lm\in\bg$.  Here $\dl_{\lm}f$
 is the infinitesimal variation under (\ref{infgt}), i.e., \be
 \dl_{\lm}f = \int_{\T} d\al\,\left( \frac{\dl f}{\dl E(\al)}\cdot
 {\rm Co}(\lm) E(\al)- \frac{\dl f}{\dl A(\al)}\cdot
 D_A\lm(\al)\right).  \ee Hence a possible choice, and the one one we
 adopt, is \be J_{\lm}(E,A)=-\la E|D_A\lm\ra, \ee where the pairing
 is, of course, between $L^2([0,2\pi],\g^*)$ and $L^2([0,2\pi],\g)$.
 This momentum map is infinitesimally equivariant in the sense that
 \be \{J_{\lm_1},J_{\lm_2}\}=-J_{[\lm_1,\lm_2]}.  \ee The charges
 $\Phi$ mentioned in the Introduction are therefore minus the
 components of the momentum map.

 An elegant way to compute the reduced space $\Sp=J^{-1}(0)/\CG$ was
given by Rajeev \ci{Raj}, and was further elaborated in \ci{Raj2}.
All results until the end of this subsection are taken from these
references; we merely add the Marsden-Weinstein reduction perspective.

Define a map $W:L^2([0,2\pi],\g)\raw C([0,2\pi],G)$ by $W(A)\equiv
W_A$, given by \be W_A(\al) =P\,\Exp\left( -\int_0^{\al}d\al'\,
A(\al')\right), \ll{wilson} \ee where $P$ denotes path-ordering, so
that $W_A(\al)$ is indeed an element of $G$; note that \be W_A(0)=e,
\ll{init} \ee so that $W$ takes values in the subspace
$C_e([0,2\pi],G)$ of functions satisfying\footnote{ In probability
theory the map $A\raw W_A$ is seen as the composition $I\circ\int_0$
of the primitive $\int_0:L^2([0,2\pi],\g)\raw C([0,2\pi],\g)$ and
Ito's map $I=P\Exp: C([0,2\pi],\g)\raw C([0,2\pi],G)$; cf.\
\ci{AHK,FRKL,MM}.} $f(0)=e$.  The path-ordered exponential is
ultimately defined as a product integral; see \ci{Dol}, and \ci{Raj2}
in the present context.  In our context, it coincides with the
solution of the differential equation\footnote{Ito's map is defined in
terms of a {\em stochastic} differential equation similar to
(\ref{W-1}) but deals with much more general function $A$, which in
our case is essentially the derivative of absolutely continuous
functions.}  \be \left(
\frac{\partial}{\partial\al}+A\right)W_A(\al)=0, \label{W-1} \ee with
initial condition (\ref{init}).  The map $W$ does not quite map
$A\in\CA$ into the gauge group, since $W_A(2\pi)$ is not necessarily
equal to $W_A(0)$.

Although elements of $S$ are not necessarily differentiable, the
constraints $J_{\lm}(E,A)=0$ for all $\lm\in\bg$ force $E$ in
$(E,A)\in J^{-1}(0)$ to have the form
\be
E(\al)={\rm Co}( W_A(\al))E, \ll{gauss}
\ee 
 where $E\in\g^*$ on the
right-hand side is constant. For abelian $G$ this simply means that
$E(\al)=E$ is independent of
$\al$. 
The expression (\ref{gauss}) implies that $(E,A)\in J^{-1}(0)$
satisfies Gauss' law
$D_AE=0$ (and {\em vice versa}).

To see the effect of passing from $J^{-1}(0)$ to $\Sp$ we look at the
cotangent bundle $T^*G$, which is canonically isomorphic to
$\g^*\times G$ \ci{AM}.  Define $\rh:J^{-1}(0)\raw T^*G$ by \be
\rh(E,A)=(E(0),W_A(2\pi)) ; \ee here $E(0)$ coincides with the $E$ on
the right-hand side of (\ref{gauss}), and $W_A(2\pi)$ is the Wilson
loop.  The adjoint action of $G$ on itself lifts to the action
$y:(\th,x)\raw ({\rm Co}(y)\th, {\rm Ad}(y)x)$ on $T^*G$.  With
respect to this lifted adjoint action, the map $\rh$ intertwines the
$\CG$-action on $S$ with the $G$-action on $T^*G$ in that $\rh\circ g=
g(0)\circ \rh$, where the $\CG$-action on the left-hand side is given
by (\ref{groupaction}). Since $\rh$ is onto, the physical phase space
is \be \Sp=(T^*G)/ {\rm Ad}(G).  \ee All physical observables that
only depend on $A$ are functions of the Wilson loop; such observables
define a certain commutative $C^*$-algebra \ci{AshIsh}. All physical
observables that polynomially depend on $E$ are expressible in terms
of the invariant elements in the universal enveloping algebra of $G$;
the simplest such element corresponds to the energy \be
h(E,A)=\frac{1}{4\pi}\int_{\T} E^2,\ll{h} \ee where the notation $E^2$
includes the trace (in the co-adjoint representation).

It goes without saying that for abelian $G$ the adjoint
action is trivial,  so that 
$\Sp=T^*G$ in that case. 
\section{Quantum reduction}
\subsection{Quantization of the unconstrained system\ll{Qu}}
We quantize the unconstrained phase space $S=T^*\CA$ by the standard
method of second quantization.  Hence we complexify the real Hilbert
space $\CA=L^2([0,2\pi],\g)$ to \be \CA_{\C}=L^2([0,2\pi],\g_{\C}),
\ee and consider the Bosonic Fock space \ci{Gui} \be
\H=\exp(\CA_{\C})=\bigoplus_{n=0}^{\infty} \otimes^n_S \CA_{\C}; \ee
here $\otimes^n_S \CA_{\C}$ denotes the symmetrized tensor product of
$n$ copies of $\CA_{\C}$.

Of special interest are the {\em coherent states} $|\sqrt{\exp}\,
A\ra$ in $\H$, defined for $|A\ra\in \CA_{\C}$ by the norm-convergent
series \ci{Gui} \be |\sqrt{\exp}\, A\ra=\sum_n (n!)^{-1/2}\otimes^n
|A\ra; \ee the notation is motivated by the property that \be
\la\sqrt{\exp}\, A |\sqrt{\exp}\, B\ra =e^{\la A|B\ra}, \ll{expip} \ee
where $\la A|B\ra$ stands for the inner product in $\CA_{\C}$.  The
importance of these vectors lies partly in the fact that one can
conveniently define a unitary representation of the gauge group $\CG$
by \be U(g)|\sqrt{\exp}\, A\ra=e^{-\half\la dg\, g^{-1}|dg\,
g^{-1}\ra+\la g^{-1} dg |A\ra}\, |\sqrt{\exp}\, gA\ra, \ll{defU} \ee
where $gA$ is defined in (\ref{GonA}).  The main term $|\sqrt{\exp}\,
({\rm Ad}(g)A-dg\, g^{-1})\ra$ illustrates that this is the second
quantization of the action (\ref{GonA}) of $\CG$ on $\CA$, the other
terms being present in order to guarantee that $U$ is a unitary group
representation. Various unitarily equivalent versions of this
representation may be found in the literature \ci{AHK,FRKL,Ism}, and
have been used in the present context \ci{dimock}; for a
three-dimensional version cf.\ \ci{LWied}.  For later use we record
the matrix element \be \la \sqrt{\exp}\, B |U(g)|\sqrt{\exp}\, A\ra=
e^{-\half\la dg\, g^{-1}|dg\, g^{-1}\ra+\la g^{-1}dg|A\ra-\la B|dg\,
g^{-1}\ra + \la B|{\rm Ad}(g)A\ra}. \ll{me} \ee

For  $f,g\in \CA_{\C}$ the usual creation- and annihilation operators $a(f)$, $a(g)^*$ satisfy the
canonical commutation relations (CCR) $[a(f),a(g)^*]=\la f|g\ra$; note that $a(f)$ is antilinear in
$f$, whereas (by implication) $a(g)^*$ is linear in $g$. The linear span is contained in the domain of
these operators, and (in the Fock representation) one has
\be
a(f)|\sqrt{\exp}\, A\ra=\la f|A\ra\,  |\sqrt{\exp}\, A\ra. \ll{af}
\ee

The linear functions $A(f)$ and $E(g)$ in $\cin(S)$ (see text after
(\ref{pb})) are quantized by \be Q(A(f))=\half(a(f)+a(f)^*) \ll{QAf}
\ee and \be Q(E(g))=-i(a(g)-a(g)^*) ,\ll{QEg} \ee respectively.  From
(\ref{pbAE}) and the CCR we see that \be
i[Q(A(f)),Q(E(g))]=Q(\{A(f),E(g)\}), \ee as desired in quantization
theory.  \subsection{Intermezzo: Wiener measure on the gauge
group}\label{interm} The subsequent construction involved in the
quantisation procedure will make use of the properties of the
(conditioned) Wiener measure $\mu_W$ on $\CG$. This measure was
constructed in \ci{AHK,FRKL,MM}, and, like the Wiener measure on
$\R^n$, is closely related to Brownian motion and the heat
equation. This relation is not very important for our purpose;
instead, the most efficient way to define $\mu_W$ is the following
method due to L. Gross \ci{LPG} (also cf.\ \ci{Kuo}).  For the theory
of promeasures and general measure theory in infinite-dimensional
spaces we refer to the reviews \ci{Kuo,DMN,AMP}; another good
reference for this subsection is section 5 of \ci{FRKL}.

Any real Hilbert space $\cal K$ has a Gaussian promeasure $\mu_c$
defined on it, which is characterized by its Fourier transform \be
\int_{\cal K}d\mu_c(\ps)e^{i\la\phv|\ps\ra} =e^{-\half Q(\phv
)}=e^{-\half\la\phv|\phv\ra},\label{Q} \ee where $Q(\phv )=\|\phv
\|^2$ is the covariance of $\mu_c$.  With this covariance, $\mu_c$ is
the canonical Gaussian measure on $\CH$; in general, any positive
quadratic form $Q$ can be the covariance.  If $\cal H$ is
finite-dimensional, $\mu_c$ is actually a measure, given by
$$ d\mu_c(x^1,\ldots ,x^n)=dx^1\ldots  dx^n e^{-\half \n x\n^2}.$$
  In general, only  so-called {\em
cylindrical functions} can be integrated  with respect to a promeasure.  
A cylindrical
function $f$ on a Hilbert space is  of the
form $f=F\circ p$, where $F$ is an integrable function on a
finite-dimensional subspace, and $p$ is the orthogonal projection
onto that subspace.  Eq.\ (\ref{Q}) provides an example: here 
 the cylindrical function is $e^{i\la\phv|\ps\ra}$. A more detailed discussion may be found in
\ci{Kuo,DMN,AMP}.

Given a measurable map $f:M\raw N$ between two measure spaces $M,N$
the image (or push-forward) of a measure $\mu$ on $M$ is the measure
$f_*\mu$ on $N$, defined by $f_*\mu(E)=\mu(f^{-1}(E))$ for all
measurable subsets $E\subset N$. In case that $M$ and $N$ are
infinite-dimensional vector spaces and $\mu$ is merely a promeasure,
this definition of $f_*\mu$ initially only applies to cylinder subsets
$E$ of $N$. It may happen that $f_*\mu$ thus defined has a countably
additive extension to the $\Sigma$-algebra generated by the cylinder
sets in $N$, so that it can be extended to a measure on $N$. But even
in that case, the volume of a non-cylindrical set $E\subset N$ must be
computed by approximating it with cylinder subsets, {\em even when
$f^{-1}(E)$ is a cylinder set in $M$}.

This comment applies to the case at hand.  In terms of the map $W$
(see (\ref{wilson})) and the promeasure $\mu_c$ on ${\cal
K}=L^2([0,2\pi],\g)$, the image $W_*\mu_c$ is initially a promeasure
on $C_e([0,2\pi],G)$, which can be extended to a measure $\nu$. The
image $E$ of $L^2([0,2\pi],\g)$ in $C_e([0,2\pi],G)$ under $W$ (which
is the subspace of continuous paths with finite energy) is not a
cylinder set, and its volume should be evaluated through the
approximation procedure mentioned above.  It then comes out that
$\nu(E)=0$, despite the fact that $\mu_c(W^{-1} (E))=1$.

Let $C_{e\raw x}([0,2\pi],G)$ be the space of continuous paths in $G$
which start at $e$ and end at $x$; we abbreviate this as $C_{e\raw
x}$.  For each $x\in G$ a measure $\mu_{x}$ on $C_{e\raw x}$ is
defined by the desintegration $\nu(A)= \int_G dx\, \mu_x(\Sigma\cap
C_{e\raw x})$, where $dx$ is the Haar measure on $G$, and $\Sigma$ is
a measurable subset of $C_e([0,2\pi],G)$.  The special case $\mu_e$ is
then a measure on the space $C_e(\T,G)$ of continuous loops in $G$
which start (and end) at $e$.  If we embed $G$ in $C(\T,G)$ as the
space of constant loops, we clearly have $C(\T,G)/C_e(\T,G)=G$ (as
groups) and $C(\T,G)=C_e(\T,G)\times G$ as measure spaces.  This
factorization finally allows us to define the Wiener measure $\mu_W$
on $C(\T,G)$ as the product $\mu_e\times \mu_H$.

Let $\ovl{\CG}= C(\T,G)$ be the space of all continuous loops in $G$;
this is the completion of $\CG$ in the supremum norm (see \ci{LPG,Kuo}
for the general theory behind such completions in measure theory).  It
is clear that $\mu_W(\ovl{\CG})=1$, whereas the comments above imply
that $\mu_W(\CG)=0$.  We summarize this discussion by

{\em The Wiener measure $\mu_W$ on the extended gauge group
$\ovl{\CG}$ is a probability measure, defined as the push-forward of
the canonical Guassian promeasure on the real Hilbert space
$L^2([0,2\pi],\g)$ by the `Wilson loop' map $W$ in (\ref{wilson}),
conditioned on the space of loops.  The gauge group $\CG$ of loops
with finite energy has volume zero w.r.t.\ the Wiener measure.  }

An important property of $\mu_W$ is its behaviour under translations;
this was first established in \ci{CM} for the original Wiener measure
on $\Bbb R^n$, and was proved in the present context of loop groups by
\ci{AHK,MM,Hsu}. It is \be d\mu_W(gh)=d\mu_W(g) \exp\left(-\half\la
dh\, h^{-1}|dh\, h^{-1}\ra-\la g^{-1} dg|dh\, h^{-1}\ra
\right),\ll{Wcov} \ee where $g\in \ovl{\CG}$ and $h\in\CG$ (the
translation property cannot be extended to all $h\in\ovl{\CG}$).

Another important property is that the measure is invariant with
respect to $g\mapsto g^{-1}$ on all $\ovl{\CG}$ \cite{MM}.  These
properties, as well as the definition of $\mu_W$, are consistent with
the heuristic formula \be
``\,d\mu_W(g)=N\prod_{\al\in\T}dg(\al)\exp\left(-\half\la
\frac{dg}{d\al}g^{-1}|\frac{dg}{d\al}g^{-1}\ra\right)\:'', \ll{heur}
\ee where $N$ is an infinite normalization constant. This formula does
not make mathematical sense, since the `Haar measure'
$\prod_{\al\in\T}dg(\al)$ on $\CG$ or $\ovl{\CG}$ does not
exist. Nonetheless, it is sometimes useful in guessing the results of
certain calculations.
\subsection{The manipulated inner product}
We now turn to the construction of the manipulated inner product
  $\mip$.  As explained in subsection \ref{discrete}, we may proceed
  in two stages, and first perform the quantum reduction with respect
  to the connected component $\CG_0$ of the identity.

In any case, we need to determine a dense domain ${\cal D}\subset
{\cal H}$ on which $\mip$ is defined; here $\H=\exp(\CA_{\C})$.  Many
different choices of ${\cal D}$ lead to the same physical Hilbert
space; a guiding principle is computational convenience. It turns out
to be appropriate to choose the following domain.

{\em The domain ${\cal D}\subset \exp(\CA_{\C})$ consists of the
finite linear span of all coherent states of the form $|\sqrt{\exp}\,
A\ra$, where $A\in\CA_{\C}$.  }

Following the proof of Prop.\ 2.2 in \ci{Gui}, one can show that
 ${\cal D}$ is dense $\exp(\CA_{\C})$.  Moreover, ${\cal D}$ is stable
 under the action of $U(g)$ for any $g\in\CG$. The advantage of this
 choice will become clear later on.

It so happens that the representation $U$ defined in (\ref{defU})
cannot be extended from $\CG$ to $\ovl{\CG}$. Nonetheless, eqs.\
(\ref{ip}), (\ref{me}), and (\ref{heur}), suggest, and almost imply,
that we should define the manipulated inner product on ${\cal D}$ by
sesquilinear extension of \be \la \sqrt{\exp}\, B |\sqrt{\exp}\,
A\ra_{\rm phys}=\int_{\ovl{\CG}} d\mu_W(g)\, e^{\la g^{-1}dg
|A\ra}e^{\la B|gA\ra}.  \ll{me0} \ee Since $g^{-1}dg$ is not
necessarily in $L^2$, the expressions $\la g^{-1}dg |A\ra$ and $\la
B|gA\ra$ should not be interpreted as inner products in $L^2$, but as
stochastic integrals \ci[\S 4.5]{Hida}.  In the present case these
stochastic integrals reduce to Stieltjes integrals (see \ci{CM} for
this remark).  We shall not dwell on this point, except by saying that
the following manipulations are all justified in the context of this
more general notion of integration.

In any case, the postulate (\ref{me0}) is jusitified by the crucial
property (\ref{cp}) (now valid on ${\cal D}$), which follows from
(\ref{defU}), (\ref{me0}) and (\ref{Wcov}).  Like the translation
formula (\ref{Wcov}), this property holds for all $h\in\CG$.  It is
important that $\cal D$ is stable under $U(\CG)$, since otherwise the
left-hand side of (\ref{cp}) would not be defined.
\section{The abelian case}
\subsection{Small gauge transformations}
We will now look at the simplest case $G=U(1)$.  First, let us reduce
with respect to the space $\CG_0$ of small gauge transformations.  In
the abelian case the product (\ref{me0}) simplifies to \be \la
\sqrt{\exp}\, B |\sqrt{\exp}\, A\ra_0=e^{\la
B|A\ra}\int_{\ovl{\CG}_0}d\mu_W(g) e^{\la g^{-1}dg|A-\ovl{B}\ra}
\label{inte}.
\ee Write $g=\exp(i\lm)$, where $\lm(2\pi)=\lm(0)$ for
$g\in\ovl{\CG}_0$.  By the definition of $\CG_0$, the set
$\{g^{-1}dg|g\in\CG_0\}$ forms the Hilbert space $P_0
^{\perp}L^2([0,2\pi], \R)$ (this would no longer be true in the
non-abelian case). Here $P_0$ is the projection onto the the constant
functions; we write $P_0 A=A_0 1$, where $A_0=(2\pi)^{-1}\int_0^{2\pi}
d\al\, A(\al)$ is the zeroth Fourier mode of $A$, and 1 is the unit
function in $L^2([0,2\pi], \R)$. The symbol $P_0^{\perp}$ will denote
the projection orthogonal to $P_0$.

Using the definition of the Wiener measure as the push-forward of the
Gaussian promeasure under the map $W$ (cf.\ (\ref{wilson}) or
(\ref{W-1})), the right-hand side of (\ref{inte}) becomes
$$
e^{\la
B|A\ra}\int_  {P_0^{\perp}L^2([0,2\pi],{\Bbb R})}{d\mu_c}(\lm)e^{\la d\lm
|A-\overline{B}\ra},
$$
where $\mu_c$ is the canonical Gaussian measure on
$P_0^{\perp}L^2([0,2\pi],{\Bbb R})$.  This step is justified because
$\ovl{\CG}_0$ is a cylinder set in $\ovl{\CG}$.

 The integral itself (without the prefactor) is computed from
(\ref{Q}), and yields $\exp[\half (A_{\perp}-\ovl{B_{\perp}})^2]$,
where $A_{\perp}=P_0^{\perp}A$ etc.  All in all, we obtain \be \la
\sqrt{\exp}\, B |\sqrt{\exp}\, A\ra_0=e^{\half(A_{\perp}^2+
\ovl{B}_{\perp}^2)} e^{2\pi \ovl{B}_0 A_0}, \ll{result} \ee where
$A^2=\la\ovl{A}|A\ra$, etc.

By definition, the induced Hilbert space $\H^0$ is the quotient ${\cal
D}/\CN$ of ${\cal D}$ by the null space $\CN$ of the manipulated inner
product, completed in the inherited norm. A trick allows us to realize
$\H^0$ in a more concrete way.

Define $V:{\cal D}\raw L^2(\R)$ by linear extension of
\begin{equation}
\la x |V|\sqrt{exp} A\ra=(\pi)^{-1/2}e^{\half A_{\perp}^2 }
\exp\left(-\frac{x^2}{2\pi}+2xA_0-\pi A_0^2\right). \ll{defV}
\end{equation}
It follows from (\ref{expip}) and a Gaussian integration that \be \la
\phv|\psi\ra_0=\la V \phv| V \ps\ra, \ll{Viso} \ee for all
$\ps,\phv\in L$; cf.\ (\ref{VIP}). Here the inner product on the
right-hand side is obviously the one in $L^2(\R)$.

This property is the whole point behind introducing the map $V$.  For
it follows that the map $V$ has the same null space $\CN$ as the
manipulated product $\mipo$, so that the quotient ${\cal D}/\CN$ is
given by the image of ${\cal D}$ under $V$.  Since the image of $V$ is
dense in $L^2(\R)$, the closure of $V{\cal D}$ is obviously
$L^2(\R)$. We may therefore identify this space with $\H^0$.

Recall the definition of a weak quantum observable; cf.\ (\ref{weak}) etc.
 Analogously to (\ref{Aphys}), the induced action $B^0$ of a weak quantum
observable $B$ on $\H^0$ is given by
\be
B^0 V|\ps\ra=VB|\ps\ra. \ll{A0V}
\ee

In the present situation notable examples of weak quantum observables,
at least with respect to the modified inner product defined by
(\ref{me0}), are $Q(A(1))$, and $Q(E(f))$ for all $f\in\CA_{\C}$; see
(\ref{QAf}), (\ref{QEg}).  The weak observability of $Q(E(f))$ is a
consequence of (\ref{me}), (\ref{me0}), and the fact that it commutes
with all gauge transformations $U(g)$. In fact, a calculation similar
to the one leading to (\ref{result}) yields \be \la \sqrt{\exp}\, B
|Q(E(f))|\sqrt{\exp}\, A\ra_{\rm phys}=-2\pi i\ovl{f}_0(A_0-\ovl{B}_0)
e^{\half(A_{\perp}^2+ \ovl{B}_{\perp}^2)} e^{2\pi \ovl{B}_0 A_0}.  \ee
Hence $Q(E(f))^0=0$ for all $f\in P_0^{\perp}\CA_{\C}$, as was to be
expected on the basis of Gauss' law.  Writing the energy (\ref{h}) as
a mode expansion $E=\sum_n E_nE_{-n}/4\pi$, this means that only the
zero mode contributes, leading to \be Q(h)^0=
\frac{1}{8\pi^2}\left((a(1)-a(1)^*)^2\right)^0.\ll{Qh} \ee

Furthermore, the Wilson loop $W_A(2\pi)$ (see (\ref{wilson})) is
quantized by \be Q(W_A(2\pi))=e^{-\half i(a(1)+a(1)^*)}.\ll{QWA} \ee
These operators are evidently constructed from $a(1)$ and $a(1)^*$.
From (\ref{A0V}) and (\ref{defV}) we obtain \be
Q(a(1))^0=x+\pi\frac{d}{dx}.  \ee Since the induction procedure
preserves the adjoint of a weak quantum observable, it follows that
\be Q(a(1)^*)^0=x-\pi\frac{d}{dx}, \ee Hence in terms of the usual
Schr\"{o}dinger position $q=x$ and momentum $p=-id/dx$ we have \bea
Q(W_A(2\pi))^0 & = & e^{-iq}; \ll{qwl}\\ Q(h)^0 & = & \half p^2
\ll{ener} \eea from (\ref{QWA}) and (\ref{Qh}), respectively.  These
are unbounded operators on $L^2(\R)$, initially defined on the linear
span of the usual coherent states, where they are essentially
self-adjoint (cf.\ \ci{RS1,RS2} for the theory of unbounded operators
on Hilbert space).
\subsection{Large gauge transformations}
Having arrived at the intermediate Hilbert space $\H^0=L^2(\R)$, we
now complete the quantum reduction by the full group $\CG$.  As
explained in subsection \ref{discrete}, the discrete group
$\pi_0(\CG)=\CG/\CG_0=2\pi \Bbb Z$ acts on $\H^0$ through a unitary
representation $U^0$. To compute this action, we write $U(n)$ for
$U(g_n)$ and note that according to (\ref{gnal}) eq.\ (\ref{defU})
specializes to \be U(n)|\sqrt{\exp}\, A\ra=e^{-\pi n^2 +2\pi
nA_0}|\sqrt{\exp}\, (A-n1)\ra. \ll{Unexp} \ee From (\ref{defV}),
(\ref{Unexp}), and (\ref{A0V}) we then infer that the corresponding
realization $U^0(n)$ on $L^2(\R)$ is simply \be \la x
|U^0(n)|\ps\ra=\la x+2\pi n|\ps\ra.\label{uz} \ee

The one-dimensional \rep s of $U_{\th}(\CG/\CG_0)$ discussed in
\ref{discrete} are here given by \be U_{\th}(n)=e^{in\th}, \ee where
$\th\in[0,2\pi)$ (note that the unitary dual $\hat{\Z}$ of $\Z$ is
$\hat{\Z}=\T$; one could consider any $\th\in\R$, and find that all
$\th$-dependent quantities are periodic in $\th$ with period $2\pi$).
We then apply (\ref{ip3}), which, with a convenient normalization
factor, now reads \be \la \phv|\ps \ra_{\rm
phys}^{\th}=2\pi\sum_{n\in\Z}e^{in\th}\int_{\R}dx\, \la\phv|x\ra\la
x+2\pi n|\ps\ra . \ll{mipbis} \ee This is well-defined on ${\cal
D}'=V{\cal D}$, which, we recall, is the linear span of all coherent
states in $L^2(\R)$. (Other domains, such as $C_c(\R)$ or the Schwartz
space ${\cal S}(\R)$ would be equally suitable, and lead to the same
result.)

One then repeats the procedure that led from $\H$ to $\H^0$.  In the
case at hand the second step of the quantum reduction procedure is
closely related to the description of the Aharonov-Bohm effect in
terms of induced representations \ci{AB}.
\subsection{Intermezzo: induced representations revisited}
More generally, whenever $\H^0$ is of the form $L^2({\sf G})$, for
some locally compact group ${\sf G}$, and $\CG/\CG_0$ is a closed
subgroup of ${\sf G}$ which acts on $\H^0$ in the right-regular
representation, the reduction from $\H^0$ to $\Hp$ is itself a special
case of the theory of induced group representations (in the sense of
Mackey; cf.\ \ci{BR}) as reformulated by Rieffel \ci{Rie}. In this
more general situation one is given a closed subgroup $H\subset {\sf
G}$ (where ${\sf G}$ and hence $H$ are assumed to be locally compact)
and a unitary \rep\ $U_{\ch}$ of $H$ in a Hilbert space $\H_{\ch}$
(here $\ch$ is some label). These data lead to a unitary \rep\
$U^{\ch}$ of ${\sf G}$ on some Hilbert space $\H^{\ch}$, said to be
induced by $U_{\ch}(H)$ \ci{BR}. As shown in \ci{Rie}, one can
construct $U^{\ch}$ and $\H^{\ch}$ as follows (also cf.\ \ci{L2}). For
simplicity we assume that ${\sf G}$ and $H$ are unimodular, so that
left- and right-Haar measures are the same; fixing a normalization, we
denote the Haar measure on ${\sf G}$ and $H$ by $dx$ and $dh$,
respectively.  This defines $L^2({\sf G})=L^2({\sf G},dx)$.  The coset
${\sf G}/H$ then has a ${\sf G}$-invariant measure $dq$, which defines
$L^2(Q)=L^2(Q,dq)$.

Choose a dense subset ${\cal D}\subset L^2({\sf G})$ as ${\cal
D}=C_c({\sf G})$, and equip $L^2({\sf G})\otimes \H_{\ch}$ with the
manipulated inner product, defined by sesquilinear extension of \be
\la\ps\ot v|\phv\ot w\ra^{\ch}_{\rm phys}=\int_H dh\, \la
v|U_{\ch}(h)|w\ra_{\ch}\int_{\sf G} dx\, \la\ps|x\ra\la
xh|\phv\ra,\ll{mipgh} \ee where $\la\,|\,\ra_{\ch}$ is the inner
product in $\H_{\ch}$. The expression (\ref{mipgh}) is well-defined on
${\cal D}\subset\ot\H_{\ch}$. Then choose a (measurable) cross-section
$s:{\sf G}/H\raw {\sf G}$, and define $V^{\ch}_s:{\cal D}\otimes
\H_{\ch}\raw L^2({\sf G}/H)\ot\H_{\ch}$ by \be \la q|V^{\ch}_s |\ps\ot
v\ra=\int_H dh\, U_{\ch}(h)|v\ra\la s(q)h|\ps\ra,\ll{Vs} \ee where
$q\in {\sf G}/H$.  A simple computation, using the invariance of $dh$
and the property \be \int_{{\sf G}/H}dq\, \int_H dh\,
f(s(q)h)=\int_{\sf G} dx\, f(x) \ee for all $f\in C_c({\sf G})$, leads
to \be \la\Psi|\Phi\ra^{\ch}_{\rm phys}=\la V^{\ch}_s\Ps\ot
v|V^{\ch}_s \Phi\ra \ee for all $\Ps,\Ph\in {\cal D}\ot\H_{\ch}$,
where the inner product on the right-hand side is in
$L^2(Q)\ot\H_{\ch}$; cf.\ (\ref{Viso}).

Therefore, by the argument that followed (\ref{Viso}), the induced
space $\Hp$ (which is the closure of ${\cal D}/\CN$) defined by
(\ref{mipgh}) may be identified with $L^2(Q)\ot\H_{\ch}$. The induced
\rep\ $U^{\ch}({\sf G})$ is then defined by the property
$U^{\ch}(x)V^{\ch}_s=V^{\ch}_sU_L$, where $U_L({\sf G})$ is the
left-regular \rep\ on $L^2({\sf G})$, tensored with the identity
acting on $\H_{\ch}$.
\subsection{The physical Hilbert space}
Comparing (\ref{mipgh}) with (\ref{mipbis}), it is clear that this
general scheme applies to the case at hand: one has ${\sf G}=\R$ and
$H=2\pi\Z$, so that ${\sf G}/H=\T=\R/2\pi\Z$, and $U_{\ch}=U_{\th}$ on
$\H_{\ch}=\C$. The Haar measure on $\Z$ is taken to be the counting
measure times $2\pi$, and the induced measure $dq$ on ${\sf G}/H$ is
just the Haar measure on $\T$. It follows that \be
\Hp=L^2(Q)\ot\H_{\ch}=L^2(\T).  \ee We choose $s:\T\raw\R$ to be
$s(\al)=\al$ for $\al\in[0,2\pi)$, upon which (\ref{Vs}) reads \be \la
\al|V^{\th}_s|\ps\ra=2\pi\sum_{n\in\Z} e^{in\th}\la\al+2\pi
n|\ps\ra. \ll{Vth} \ee

The condition for an operator $B$ on $L^2(\R)$ to induce a
well-defined physical operator $B^{\rm phys}$ on $L^2(\T)$ is
(\ref{weak}), with $\la\,|\,\ra_{\rm phys}$ replaced by
$\la\,|\,\ra_{\rm phys}^{\th}$, given by (\ref{mipbis}). Explicitly,
this condition is equivalent to \be \int_{\R}dx\, \la\phv|x\ra\la
x+2\pi n|B|\ps\ra= \int_{\R}dx\, \la\phv|B|x\ra\la x+2\pi n|\ps\ra
\ll{condition} \ee for all $\ps,\phv\in L$.  The physical observable
$B^{\rm phys}$ is then given by \be B^{\rm phys}V^{\th}_s=V^{\th}_s B
\ll{Bphys}.  \ee

Condition (\ref{condition}) is satisifed by all differential operators
with constant coefficients, such as $p$ and $p^2$, but not by the
position operator $q$. Instead, one must consider periodic functions
of $x$ with period $2\pi$ (acting on $L^2(\R)$ as multiplication
operators). The quantization of the Wilson loop (\ref{qwl}) is a case
in point. From (\ref{Bphys}) and (\ref{Vth}) we then obtain \be \la
\al|Q(W_A(2\pi))^{\rm phys}|\ps\ra =e^{-i\al}\la \al| \ps\ra.  \ee For
any power of the Schr\"{o}dinger momentum $p$ (such as the energy
(\ref{ener})) we find the formal expression \be \la\al|(p^n)^{\rm
phys}|\ps\ra=\left(-i\frac{d}{d\al}\right)^n\la\al |\ps\ra.  \ee This
brings us to a crucial aspect of our technique, namely the fact that
our method of constructing $\Hp$ automatically selects a domain of
definition for unbounded weak quantum observables. This domain is the
image under $V$ (or, in the present case, $V^{\th}_s$) of the original
domain (assuming the latter to be contained in the domain ${\cal D}$
of the manipulated inner product).  In the present case the $p^n$ were
initially defined on the domain ${\cal D}'$ (i.e., the linear span of
all coherent states in $L^2(\R)$). One easily verifies that the
closure of $p^n$ coincides with the closure of $p^n$ defined on the
domain $\cin(\R)\cap L^2(\R)$. It then follows e.g.\ from Theorem
11.2.3 in \ci{BR} that $p^n$ is essentially self-adjoint on ${\cal
D}'$ for all $n$.

The image ${\cal D}_{\th}$ of ${\cal D}'$ under $V^{\th}_s$ is the
domain ${\cal D}_{\th}$, consisting of the smooth functions
$\ps\in\cin([0,2\pi])$ for which all derivatives $\ps^{(n)}$,
$n=0,\ldots$ satisfy the twisted boundary conditions
$\ps^{(n)}(2\pi)=\exp(-i\th)\ps^{(n)}(0)$.  As in Example X.1.1 in
\ci{RS2} one verifies that $(p^n)^{\rm phys}$ is essentially
self-adjoint on ${\cal D}_{\th}$ for all $n$. This particularly
applies to the enery (\ref{ener}), where $n=2$. Hence one obtains a
uniquely determined observable $Q(h)^{\rm phys}$, defined as the
self-adjoint closure of $\half p^2$ (initially defined on ${\cal
D}_{\th}$).  Its eigenfunctions $\ps_n$, $n\in\Z$, are
$\la\al|\ps_n\ra=\exp(i\al(n-\th/2\pi))$, with eigenvalues
$E_n=\half(n-\th/2\pi)^2$. This is one way of seeing how the
$\th$-parameter enters the physical theory.

\end{document}